\documentclass[useAMS,usenatbib]{mn2e}

\usepackage[pdftex]{graphicx}
\usepackage[utf8]{inputenc}
\usepackage{amsmath}
\usepackage{amssymb}
\usepackage{color}
\usepackage{tabularx}
\usepackage{multirow}
\usepackage{sidecap}
\usepackage{array}
\newcolumntype{Z}{>{\centering\let\newline\\\arraybackslash\hspace{0pt}}X}

\newcommand{\Msun}{\ensuremath{\text{M}_\odot}}
\newcommand{\PhiN}{\phi}
\newcommand{\Phiph}{\Phi_\text{ph}}
\renewcommand{\d}{{\text{d}}}
\newcommand{\grad}{{\bf \nabla}}

\newcommand{\mnras}{MNRAS}
\newcommand{\apj}{ApJ}

\newcommand{\aap}{A\&A}
\newcommand{\nat}{Nature}
\newcommand{\pasa}{PASA}
\newcommand{\pasj}{PASJ}

\newcommand{\prd}{Phys. Rev. D}
\newcommand{\physrep}{Physics Reports}
\newcommand{\aj}{AJ}

\pagerange{\pageref{firstpage}--\pageref{lastpage}} \pubyear{2013}

\title{Polar ring galaxies as tests of gravity}

\author[F. L\"ughausen et al.]
{F.~L\"ughausen$^1$\thanks{fabian@astro.uni-bonn.de} ,
B.~Famaey$^2$,
P.~Kroupa$^1$,
G. Angus$^3$,
F. Combes$^4$,
G. Gentile$^{5,6}$,
\newauthor 
O. Tiret$^7$,
H. Zhao$^8$
\\
$^1$Argelander-Institut f\"ur Astronomie, Auf dem H\"ugel 71, D-53121 Bonn, Germany\\
$^2$Observatoire Astronomique, Universit\'e de Strasbourg, CNRS UMR 7550, France\\
$^3$Astrophysics, Cosmology \& Gravity Centre, University of Cape Town, South Africa\\
$^4$Observatoire de Paris, LERMA, France\\
$^5$Sterrenkundig Observatorium, Ghent University, Belgium\\
$^6$Department of Physics and Astrophysics, Vrije Universiteit Brussel, Pleinlaan 2, 1050 Brussels, Belgium\\
$^7$Laboratoire d'Astrophysique, Ecole Polytechnique F\'ed\'erale de Lausanne, Switzerland\\
$^8$School of Physics \& Astronomy, University of St Andrews, Scotland, UK}

\begin{document}

\date{\today}
\maketitle

\begin{abstract}
Polar ring galaxies are ideal objects with which to study the three-dimensional shapes of galactic gravitational potentials since two rotation curves can be measured in two perpendicular planes. Observational studies have uncovered systematically larger rotation velocities in the extended polar rings than in the associated host galaxies. In the dark matter context, this can only be explained through dark halos that are systematically flattened along the polar rings. Here, we point out that these objects can also be used as very effective tests of gravity theories, such as those based on Milgromian dynamics (MOND). We run a set of polar ring models using both Milgromian and Newtonian dynamics to predict the expected shapes of the rotation curves in both planes, varying the total mass of the system, the mass of the ring with respect to the host, as well as the size of the hole at the center of the ring. 
We find that Milgromian dynamics not only naturally leads to rotation velocities being typically higher in the extended polar rings than in the hosts, as would be the case in Newtonian dynamics without dark matter, but that it also gets the shape and amplitude of velocities correct. Milgromian dynamics thus adequately explains this particular property of polar ring galaxies.
\end{abstract}


\section{Introduction}
Assuming General Relativity to be the correct description of gravity at all scales, data ranging from the largest scales (e.g., the Cosmic Microwave Background) to galactic scales can be interpreted as a Universe dominated by dark energy and dark matter. The nature of these is among the most challenging problems of modern physics. While dark energy is generally assumed to be a non-vanishing vacuum energy represented by a cosmological constant $\Lambda$ in Einstein's equations, the currently most favored dark matter candidates are neutral fermionic particles, which condensed from the thermal bath of the early Universe \citep{Bertone2005,Strigari2012}, known as ``cold dark matter'' (CDM) particles. 

On galaxy scales, predictions of this concordance cosmological model ($\Lambda$CDM) are difficult to reconcile with observations~\citep{Disney08, Peebles, Kroupa1, Kroupa2, KroupaPawlowskiMilgrom2013}. For instance, many observed scaling relations~\citep[see][for a review]{FamMcgaugh} involve the universal appearance of an acceleration constant $a_0 
	\approx \Lambda^{1/2} 
	\approx 10^{-10}{\rm m} \, {\rm s}^{-2} 
	\approx 3.6\,\text{pc}\,\text{Myr}^{-2}
$, whose origin is unknown in the standard context. 
For instance, this constant defines the zero-point of the Tully-Fisher relation, the transition of the acceleration at which the mass discrepancy between baryonic and dynamical mass appears in the standard picture, the transition central acceleration between dark-matter dominated and baryons-dominated galaxies (within Newtonian gravity), and it defines a critical mean surface density for disk stability \citep{FamMcgaugh}. These independent occurences of $a_0$ are not at all understood in the standard context, whereas, surprisingly, all these relations can be summarized by the empirical formula of \citet{Mil83}. For this formula to fit galaxy rotation curves, the above quoted value of $a_0$ can vary only between $0.9$ and $1.5 \times 10^{-10}{\rm m} \, {\rm s}^{-2}$,  but once a value is chosen, all galaxy rotation curves must be fitted with a single value \citep{things}. We choose here the median value $a_0=1.2 \times 10^{-10}{\rm m} \, {\rm s}^{-2}$, as per \citet{things}.

The success of Milgrom's empirical formula lends weight to the idea that the gravitational field in galaxies can be described by Milgromian Dynamics (also known as Modified Newtonian Dynamics or MOND). 
Milgromian dynamics naturally explains the intimate relation between the distribution of baryons and the gravitational field in galaxies, and explains all the aforementioned occurrences of $a_0$ in galactic dynamics without any fine-tuning. Given the predictive nature of Milgromian dynamics on galaxy scales,
it is of great interest to test whether the formula can explain all probes of galactic gravitational potentials, beyond spherical and axisymmetric systems where it has mostly been tested up to now.

Polar ring galaxies (PRGs) are non-axisymmetric systems featuring an outer ring of stars and gas rotating over the poles. The host galaxy is usually characterized by a compact bulge and a small bright gas-poor disk, while the gas-rich polar structure has photometric properties roughly similar to those of gas-rich spirals \citep[e.g.,][]{Whitmore1990}. 
The observer can typically measure two perpendicular rotation curves
\citep{Schweizer1983,Sackett1990,Sackett1994,Reshetnikov1994,CombesArnaboldi,Iodice03,Iodice2006,Iodice2}, 
one in the host, often by deriving an asymmetric-drift corrected rotation curve from the observed stellar kinematics \citep[see, e.g.,][]{CombesArnaboldi}, and one in the polar ring, by directly measuring the velocity of the HI gas. This makes PRGs ideal test objects for gravity theories, because any given theory of gravity then has to explain two rotation curves in two perpendicular planes, both derived from the same baryonic mass density distribution. Interestingly, observational studies \citep{Iodice03, Moiseev11} consistently show rotational velocities in the polar rings to be systematically larger than in the hosts. 
These observations may only be explained in the standard context by dark halos systematically flattened along the polar rings \citep[see][]{Iodice03}. In any case, given these specific observational properties of PRGs, it is of great interest to investigate whether the general predictions of Milgromian dynamics for such objects would conform with these observational properties, 
namely whether larger velocities in the extended polar rings than in the hosts are a generic prediction of Milgromian dynamics, by exploring a wide range of baryonic mass distributions.

In Sect.~\ref{sect:mond}, we recall the basics of Milgromian dynamics and the specific quasi-linear formulation we are dealing with. We then present a grid-based prescription to solve the modified Poisson equation (Sect.~\ref{sect:pdm_calculation}) and an iterative method to find rotation curves of non-circular orbits (Sect.~\ref{sect:rotcurves}), and apply it to a set of models in Sect.~\ref{sect:models}. Results are presented and discussed in Sect.~\ref{sect:results} and we conclude in Sect.~\ref{sect:conclusions}.

\section{Milgromian dynamics}\label{sect:mond}
In recent years, a plethora of generally covariant modified gravity theories have been developed, yielding a Milgromian behavior in the weak-field limit \citep{FamMcgaugh}.  One such recent formulation \citep{BIMOND} has a non-relativistic quasi-static weak field limit, for a specific given set of parameters, yielding the following Poisson equation:
\begin{equation}
\nabla^2 \Phi=4 \pi G \rho_\text{b} + \nabla \cdot \left[ \nu\left(|{\mathbf\grad} \PhiN|/a_0\right) {\mathbf\grad} \PhiN \right],
\label{pois}
\end{equation}
where $\Phi$ is the total (Milgromian) potential, 
$\rho_\text{b}$ is the baryonic density, 
$\PhiN$ the Newtonian potential such that $\nabla^2 \PhiN=4 \pi G \rho_\text{b}$, and where $\nu(x) \rightarrow 0$ for $x \gg 1$ and $\nu(x) \rightarrow x^{-1/2}$ for $x \ll 1$.
One family of functions that fulfills the definition of $\nu(x)$ \citep[see, e.g.][]{FamMcgaugh} is
\begin{equation}
\nu(x)=\left[{1+(1+4x^{-n})^{1/2} / 2}   \right]^{1/n} -1 .
\label{eq:nu}
\end{equation}
Hereafter, when not stated otherwise, we use $n=1$, a function which is known to reproduce well the rotation curves of most spiral galaxies\footnote{Nowadays, galaxy data still allow some, but not much, wiggle room on choosing the interpolating function $\nu(x)$ (Eq.~\ref{eq:nu}): they tend to favor the $n=1$ function from the family used here; some interpolation between $n=1$ and $n=2$; or functions from other families which actually reduce, for accelerations typical of galaxies, to the $n=1$ case used here. See, e.g., Sect.~6.2. of \citet{FamMcgaugh} for a review.} \citep{things}.

This means that the total gravitational potential $\Phi = \PhiN + \Phiph$ can be divided into a classical (Newtonian) part, $\PhiN$, and a Milgromian part, $\Phiph$.
The matter density distribution $\rho_\text{ph}$ that would, in Newtonian gravity, yield the additional potential $\Phiph$, and therefore obeys $\nabla^2 \Phiph = 4\pi G \rho_\text{ph}$, is known in the Milgromian context as the ``phantom dark matter'' (PDM)\footnote{In the Milgromian context, PDM is not real matter but a numerical ansatz which helps to compute the additional gravity predicted by Milgromian dynamics and gives it an analog in Newtonian dynamics.} density,
\begin{equation}
	\rho_\text{ph} = \frac{\nabla \cdot \left[ \nu\left(|{\mathbf\grad} \PhiN|/a_0\right) {\mathbf\grad} \PhiN \right]}{4\pi G} \,.
\label{rho_phantom}
\end{equation}
This is the density of dark matter that would boost the Newtonian gravitational field to give precisely the same effect as the boost of gravity predicted by Milgromian dynamics. For a disk galaxy, it will typically resemble a round isothermal halo at large radii, but exhibits an additional disk of phantom dark matter aligned with the baryonic disk (but with a different scale length and height) most prominent at smaller radii \citep{mondhalo}.
At each spatial point, $\rho_\text{ph}$ is a non-linear function of the Newtonian potential.
As the non-linearity of Eq.~\ref{pois} is only present on the right-hand side of the equation, it is called the quasi-linear version of Modified Newtonian dynamics \citep[][]{QUMOND} whereas in older versions of Milgromian dynamics theories the Laplacian operator on the left-hand side was replaced by a non-linear one \citep{BM84}. 
In the next section, we present a grid-based prescription to calculate the PDM density. We will then be able to compute the rotation curves from the velocity of the closed orbits crossing the planes of symmetry of the non-axisymmetric system (i.e. the plane of the host galaxy and of the polar ring). 

\section{Grid-based calculation of the phantom dark matter density}\label{sect:pdm_calculation}
The PDM density that would source the Milgromian force field in Newtonian gravity
 is defined by Eq.~\ref{rho_phantom} and can be calculated from the known classical (Newtonian) potential $\PhiN$.
To evaluate this term, we devise a numerical, grid-based scheme that calculates $\rho_\text{ph}$ from any (discrete) Newtonian potential $\PhiN^\text{i,j,k}$ 
(see \citealp[][Eq.~35]{FamMcgaugh}; \citealp{Angus1,Angus2}).

\begin{figure}
   \centering
   \includegraphics[width=7cm]{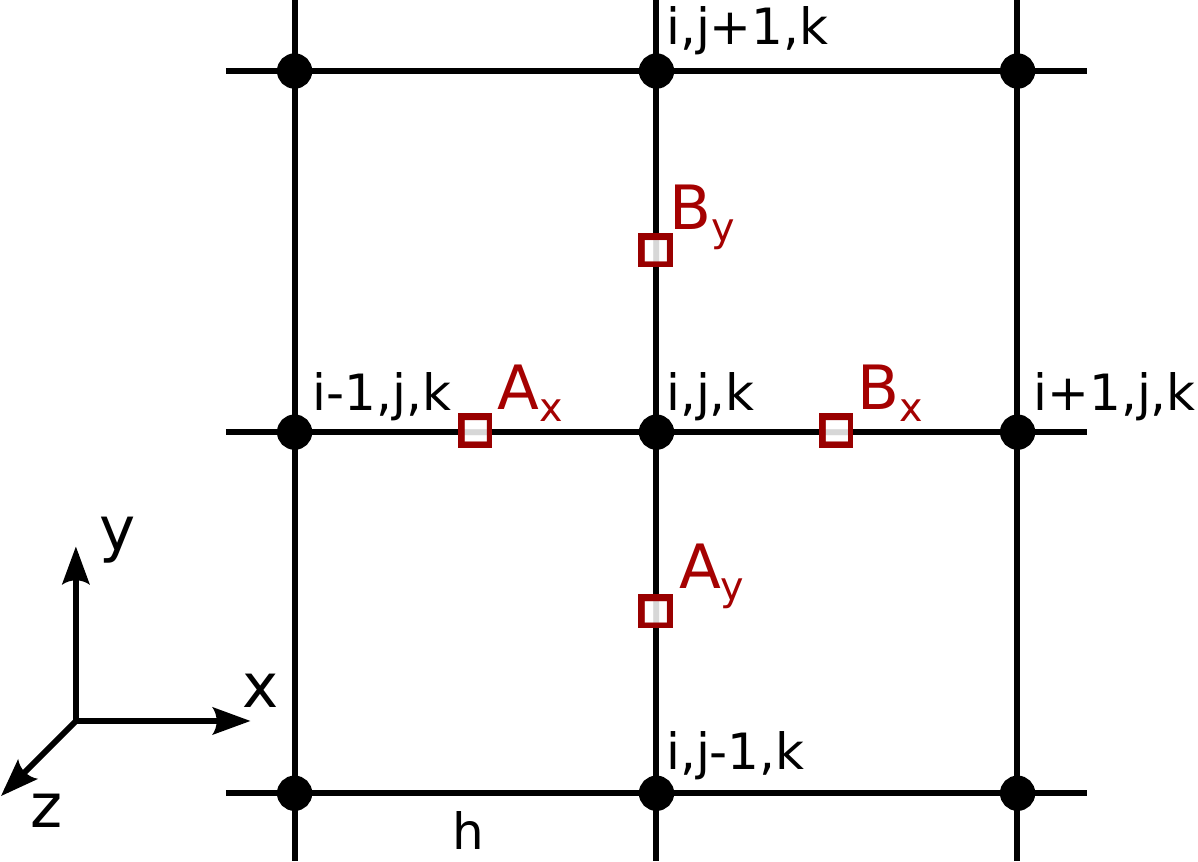}
      \caption{Illustration of the discretisation scheme in the x-y plane, referring to Eq.~\ref{eq:phantom_dm}. The grid has $N^{3}$ nodes $\left(\text{i},\text{j},\text{k}\right)$ with $\text{i},\text{j},\text{k} \in \lbrace 1\dots N\rbrace$ that are separated by a constant grid step size $h$. The values of $\nu(x)$ are evaluated at the points $\text{A}_{k}$ and $\text{B}_{k}$ in $k$-direction using Eq.~\ref{eq:gradient}.}
\label{fig:grid}
\end{figure}

The discrete form of Eq.~\ref{rho_phantom} then reads
\begin{align}
	\rho_\text{ph}^\text{i,j,k} = \frac{1}{4\pi G}  \frac{1}{h^{2}}  \left[\right.\ 
		&\left( \PhiN^{\text{i+1,j,k}} - \PhiN^\text{i,j,k} \right) \nu_{\text{Bx}} \label{eq:phantom_dm}\\
	\ -	& \left( \PhiN^\text{i,j,k} 	- \PhiN^\text{i-1,j,k}	\right) \nu_{\text{Ax}} \nonumber \\
	\ +	& \left( \PhiN^\text{i,j+1,k}	- \PhiN^\text{i,j,k}	\right) \nu_{\text{By}} \nonumber \\
	\ -	& \left( \PhiN^\text{i,j,k}	- \PhiN^\text{i,j-1,k}	\right) \nu_{\text{Ay}} \nonumber \\
	\ +	& \left( \PhiN^\text{i,j,k+1}	- \PhiN^\text{i,j,k}	\right) \nu_{\text{Bz}} \nonumber \\
	\ -	& \left( \PhiN^\text{i,j,k}	- \PhiN^\text{i,j,k-1}	\right) \nu_{\text{Az}} \left.\right] \,,\nonumber
\end{align} 
with $h$ being the constant one-dimensional grid step size. The whole discretization scheme is illustrated in Fig.~\ref{fig:grid}. Note that this equation was first derived for older versions of Milgromian dynamics by \citet{BradaMilgrom1999} and \citet{Tiret2007} and we derive it for the quasi-linear formulation here \citep[see also][]{Angus1,Angus2}.
The function $\nu(x)$ is evaluated at the points marked by squares in Fig.~\ref{fig:grid}.
The gradient of $\PhiN$ in $\nu\left(|{\mathbf\grad} \PhiN|/a_0\right)$ at the point $\text{B}_\text{x}$ has been approximated by
\begin{align}\label{eq:gradient}
\grad \PhiN = \frac{1}{4h}\begin{pmatrix} 
	4\left(	\PhiN^\text{i+1,j,k}	- \PhiN^\text{i,j,k} \right) \\
	\PhiN^\text{i+1,j+1,k}- \PhiN^\text{i+1,j-1,k} + \PhiN^\text{i,j+1,k} - \PhiN^\text{i,j-1,k} \\
	\PhiN^\text{i,j,k+1}	- \PhiN^\text{i,j,k-1} + \PhiN^\text{i+1,j,k+1} - \PhiN^\text{i+1,j,k-1} \\
\end{pmatrix} \,.
\end{align}
Having evaluated this field, the Newtonian Poisson equation,
$ \nabla^2 \Phi=4 \pi G \left( \rho_\text{b} + \rho_\text{ph} \right)$,
can be solved to find the effective Milgromian force field. This can be done using the same grid. In the present paper, grids with a resolution of $h=0.23\,\text{kpc}$ are used to calculate the rotation curves (see next section) from 2~to~16\,kpc, and grids with $h=0.47\,\text{kpc}$ for radii larger than 16\,kpc.
The resolution was chosen such that it is sufficiently fine that the form of the rotation curves does not change if the resolution is further increased.

\section{Calculation of rotation curves}\label{sect:rotcurves}
\begin{figure}
   \centering
   \includegraphics[width=7cm]{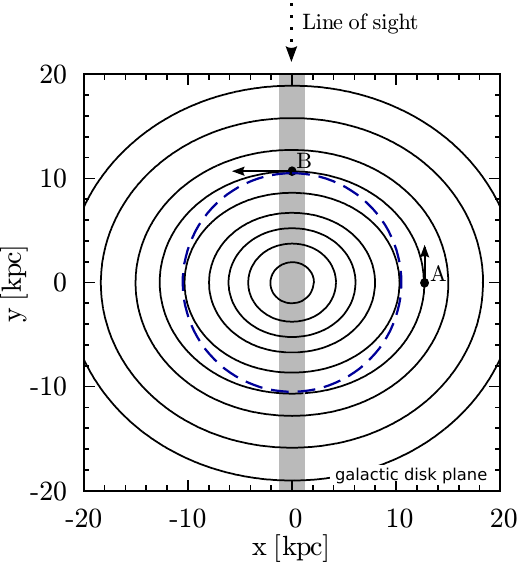}
      \caption{Closed orbits (black, solid lines) within the potential of the benchmark model (Seq.~1, $M_\text{PR} = 1.45\,M_\text{disk}$) in the plane of the host galaxy. The polar ring, which is located in the $y$-$z$ plane, is illustrated by the thick, grey line. The blue, dashed line is a circle which, by comparison, demonstrates the non-circularity of the closed-loop orbits.
      The major axis of the eccentric orbits points in the $x$-direction, because the test particles orbiting in the galactic disk ($x$-$y$) plane ``fall" through the polar ring ($y$-$z$) plane, i.e. they feel a stronger acceleration in $x$ than in $y$ direction. However, to fulfill closed orbits, the oscillation period in both directions must be the same, which means that the oscillation amplitude in the $x$ direction must be larger (major axis) than that in the $y$ direction (minor axis).
      The rotation velocity thus is minimal at point A (along the line of sight) and maximal at point B.
      }
\label{fig:orbits}
\end{figure}

From a given Milgromian potential $\Phi$ (see next section for detailed description of PRG models), rotation curves (or rather, their non-circular equivalent in non-axisymmetric configurations) can be calculated. In an axisymmetric potential, the circular rotation velocity $v(r)$, which results in closed orbits with radius $r$, readily follows as $v(r) = \sqrt{-r \cdot\d \Phi / dr}$ in the plane of the galactic disk. This equation however loses validity in a non-axisymmetric potential like the one of a PRG, because the closed orbits are generally not circular. 
The existence of two massive systems in perpendicular orientations means that circular orbits do not exist in either system, neither the equatorial nor the polar one: in each plane, the potential well corresponding to the other perpendicular system produces the equivalent of a (non-rotating) bar along the line of nodes.
In that case, it becomes necessary to obtain the velocities in the disk and polar ring in a more general way. In the present work, an iterative method is applied: test stars are shot through the galactic potential, which is computed numerically from analytical density distributions following the prescription in Sect.~\ref{sect:pdm_calculation}. The initial velocity (perpendicular to the radius) of these test particles is adjusted until a closed orbit is found. The orbit is integrated using the simple leapfrog integration scheme. Typical closed orbits in a PRG potential are shown in Fig.~\ref{fig:orbits} (detailed description of the model in the next section).

\section{Models}\label{sect:models}

\begin{figure}
   \centering
   \includegraphics[width=8.2cm]{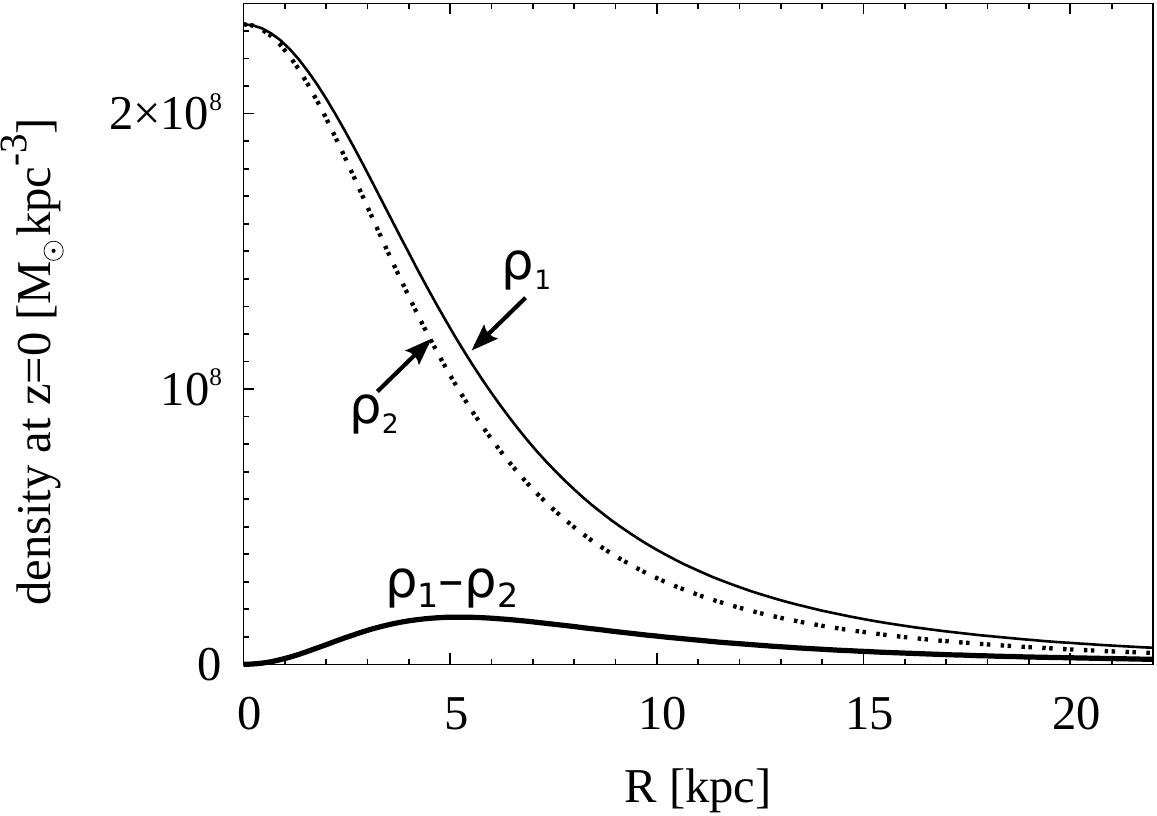}
      \caption{The polar ring density profile is made from the difference of two Miyamoto-Nagai density profiles ($\rho_1$ and $\rho_2$). Its density is zero at the centre. This example is a demonstration of a ring with a total mass $M=9.5\times 10^9\,\Msun$, a scale height $h_z=0.3\,\text{kpc}$ and scale radii $h_{\text{r}_{1}} = 6.8\,\text{kpc}$ and $h_{\text{r}_{2}} = 5.95\,\text{kpc}$ }
\label{fig:ringmodel}
\end{figure}

In order to explore the consequences of Milgromian dynamics for the rotation curves in polar rings, we start from a benchmark model adopted from \citet{CombesArnaboldi}, which represents a prototypical example of polar ring galaxy (NGC~4650A). From this model, we will construct a Milgromian potential in which the orbits of test particles will be computed. The host galaxy is made of a small Plummer bulge \citep{plummer} weighing $M_\text{b}=0.2 \times 10^9\,\Msun$, with a Plummer radius $r_\text{p}=0.17\,\text{kpc}$, 
\begin{equation}
\rho_\text{b} (r )= \left( \frac{3 M_\text{b}}{4 \pi r_\text{p}^3} \right) \left( 1 + \left(\frac{r}{r_\text{p}}\right)^2 \right)^{-5/2},
\end{equation}
and of a Miyamoto-Nagai disk \citep{MN} with disk mass $M_d=11 \times 10^{9}\,\Msun$, scale-length $h_r=0.748\,\text{kpc}$ and scale-height $h_z=0.3\,\text{kpc}$,
\begin{equation}
\begin{array}{l}
\rho_\text{d}(R,z)=\left(\frac{{h_\text{z}}^2 M_\text{d}}{4 \pi}\right) \times \\
\frac{h_\text{r} R^2+(h_\text{r}+3\sqrt{z^2+{h_\text{z}}^2})\left(h_\text{r}+\sqrt{z^2+{h_\text{z}}^2}\right)^2 }
{ \left[h_\text{r}^2+\left(h_\text{r}+\sqrt{z^2+{h_\text{z}}^2}\right)^2\right]^{5/2}\left(z^2+{h_\text{z}}^2\right)^{3/2} }.
\end{array}
\end{equation}
To this parent galaxy a polar ring of stars and another one of gas is added\footnote{We do not consider the possibility of two gaseous disks to avoid orbits crossing, unless there is a very small gas disk and a much larger gaseous polar ring which never intersect. In our models, we make no distinction between gas and stars in the host disk.}. Each ring is built by the difference of two Miyamoto-Nagai density distributions of scale height $h_\text{z} = 0.3\,\text{kpc}$, and with scale radii $h_{\text{r}_{1}}$ and $h_{\text{r}_{2}}$ (see Fig.~\ref{fig:ringmodel} for an illustration, where one sees that the baryonic density is precisely zero at the center and positive elsewhere). The masses of these two disks are chosen such that their difference equals the total mass of the ring and that the central mass density of the ring is zero.
The stellar ring weighs $9.5 \times 10^{9}\,\Msun$, and has $h^\text{st}_{\text{r}_{1}} = 6.8\,\text{kpc}$ and $h^\text{st}_{\text{r}_{2}} = 5.95\,\text{kpc}$, while the gaseous ring weighs $6.4 \times 10^{9}\,\Msun$, and has $h^\text{gas}_{\text{r}_{1}} = 15.3\,\text{kpc}$ and $h^\text{gas}_{\text{r}_{2}} = 3.4\,\text{kpc}$.
\begin{figure}
   \centering
   \includegraphics[width=9cm]{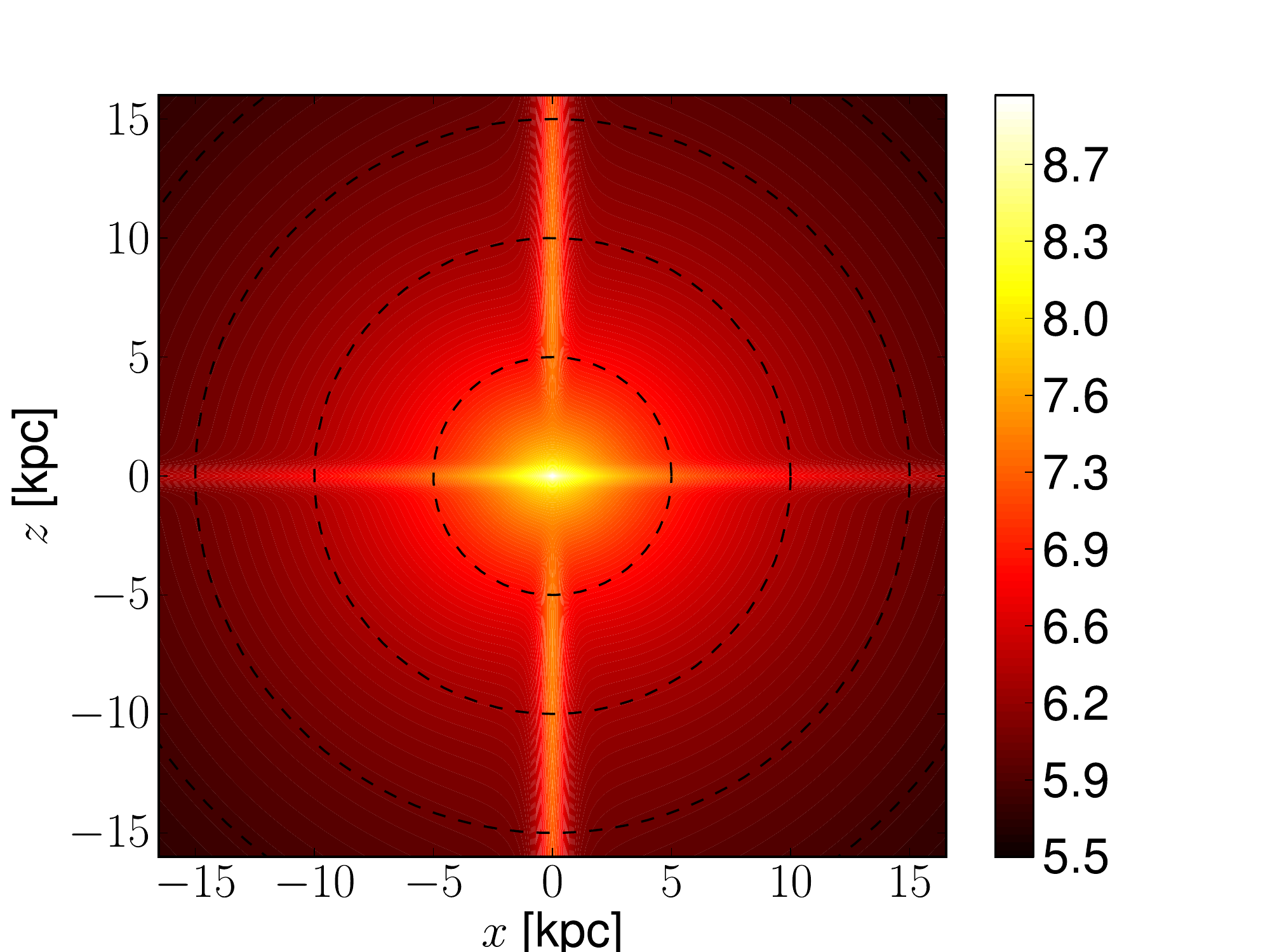}
      \caption{The figure shows contours of the mass density of the phantom dark matter of the benchmark model seen edge-on ($x$-$z$ plane, $y=0$). 
      The colour bar shows the logarithm of the PDM density in $\Msun\,\text{kpc}^{-3}$.
	In addition to the oblate phantom dark matter (PDM) halo, the horizontal overdensity shows clearly the PDM in the host disk and the vertical over density shows the PDM in the polar ring.
	The dashed circles are to demonstrate that the phantom halo is nearly spherical and slightly oblate.}
\label{fig:phantom}
\end{figure}
The parameters of these rings are such that the density in the center is zero and positive everywhere else. Note that these parameters are adopted exactly as per \citet{CombesArnaboldi}, but that in reality, some freedom on the mass of the stellar components in both the host and ring of NGC~4650A is possible. As we do not intend here to make a full detailed fit of the rotation curves of NGC~4650A, which will be the topic of a following paper, including other individual polar ring systems observed in HI with the WSRT (Westerbork Synthesis Radio Telescope), we keep the benchmark model as such. From this density of baryonic matter, we compute the corresponding PDM density using Eq.~\ref{eq:phantom_dm}. The computed distribution of PDM in the plane orthogonal to both the disk and ring is plotted in Fig.~\ref{fig:phantom}. This figure illustrates that, in addition to the oblate PDM halo (isothermal at large radii), there are also two PDM disks aligned with the baryonic disks of the host and of the polar ring.

To investigate whether the results we obtain (see Sect.~\ref{sect:results}) for this benchmark model are actually a generic prediction of Milgromian dynamics, we will then vary the parameters of this benchmark model in 5 different ways (including changing the ring into a Miyamoto-Nagai disk) computing a total of 45 models spanning a wide range of parameters. All models and their different parameters are summarized in Table~1.

\begin{itemize}
\item First, the density of the polar rings and accordingly their mass, $M_\text{PR}$, relative to the mass of the host galaxy is varied. The resulting models are collected into two sequences: Sequence~1 and Sequence~2. Starting from $M_\text{PR} = 0$, the ring mass is increased to $M_\text{PR} = \left( 0.1, 0.25, 0.33, 0.5, 0.75, 1~\text{and}~1.45\right) \times M_\text{disk}$. Sequence~1 has a constant disk mass of $M_\text{disk} = 11\times 10^9\,\Msun$, Sequence~2 has $M_\text{disk} = 33\times 10^9\,\Msun$.

\item To obtain Sequence~3, Sequence~1 is repeated while replacing the polar ring by a polar disk of mass $M_\text{PD}$. 
This polar disk is shaped like a Miyamoto-Nagai density distribution with $h_\text{r} = 0.748\,\text{kpc}$ and $h_\text{z} = 0.3\,\text{kpc}$. The polar disk mass, $M_\text{PD}$, is varied analogously to Sequence~1 and~2. \\
In this sequence, the model with $M_\text{PD}/M_\text{disk} = 1$ is symmetric, the rotation curves in the galactic plane and in the polar plane are consequently identical.

\item To investigate the influence of the polar ring shape, we vary the shape of the ring in Sequence~4. The size parameters of the gaseous ring ($h_{\text{r}_1}^\text{gas}$ and $h_{\text{r}_2}^\text{gas}$) and of the stellar ring ($h_{\text{r}_1}^\text{st}$ and $h_{\text{r}_2}^\text{st}$) are summarized in Table~1. 

\item Models of Sequence~5 have again the same structural parameters as the benchmark model, but their densities are scaled such that their total masses (of the whole system) range from $6.8\times 10^9\,\Msun$ to $10^{11}\,\Msun$. The size parameters remain unchanged. 

\item Sequence~6 is a series of four different Milgromian potentials computed for the benchmark model for $n=1,2,3,4$ in Eq.~\ref{eq:nu}, to check whether the qualitative results are independent of the $\nu$-function.
\end{itemize}

\begin{figure}
   \centering
   \includegraphics[width=8.5cm]{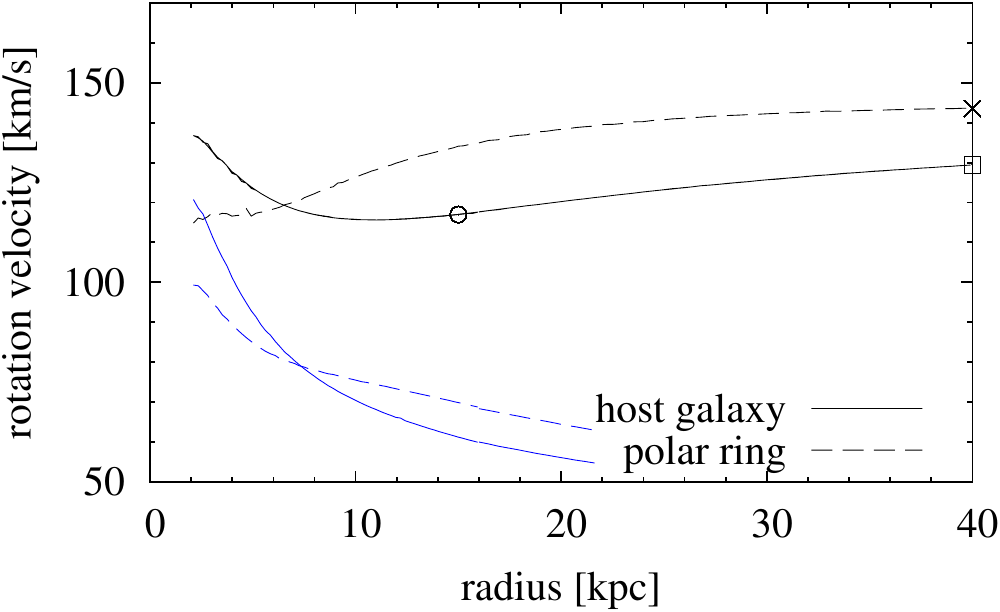}
      \caption{Rotation curves of the host galaxy (black, solid line) and polar ring (black, dashed line) for the benchmark model. These rotation curves are derived from the Milgromian potential. 
      For comparison, the blue lower lines show the rotation curves derived from the Newtonian potential of the same model.
      Observationally, the rotation velocity in the host is generally obtained indirectly from the measured stellar velocity dispersion, which means that the maximum velocity is most likely not measured in the very flat part. To account for this issue, the theoretical rotation velocity from the models are computed at both $r=40\,\text{kpc}$ and $r=15\,\text{kpc}$.
      The circle and cross at these two radii thus mark the rotation velocities that are summarized in Fig.~\ref{fig:tf}. In the case where both the host and PR are gas-rich, the PR curve should start where the host curve ends (e.g., at 15 \,\text{kpc}) to avoid collisional orbit crossing.}
\label{fig:rotcurves}
\end{figure}

\section{Results}\label{sect:results}
The rotation curves of the benchmark model are presented in Fig.~\ref{fig:rotcurves}, both in Newtonian without dark matter and Milgromian dynamics. In both cases, the velocity is higher in the ring, but of course, Milgromian dynamics is needed to get the amplitude and shape of the rotation curves right. This is a key result for Milgromian dynamics, since observational studies \citep{Iodice03} have measured rotation velocities in extended polar rings being systematically larger than in the host galaxies when both systems are seen roughly edge-on. If this result is generic, it means that this particular property of polar ring galaxies does not exclude Milgromian dynamics. To investigate whether this is indeed a generic prediction of Milgromian dynamics, we vary the parameters of this benchmark model as explained in Sect.~\ref{sect:models}. Altogether, the rotation velocities, corresponding to the line-of-sight velocities that would be measured when observing the PRG edge-on, of 45 models in total are evaluated. 
The rotation curves of the benchmark model are presented in Fig.~\ref{fig:rotcurves}. 
Fig.~\ref{fig:multiplot} contains all evaluated rotation curves of Sequences~1-4. See the caption for more details.
Sequence~6 is plotted in Fig.~\ref{fig:seq_nu}.
Finally, the shape of the rotation curves of Seq.~5 are all similar to the benchmark model, but their amplitude varies as in Fig.~\ref{fig:tf}.

\subsection{Newtonian dynamics}
Fig.~\ref{fig:rotcurves} and the first column of Fig.~\ref{fig:multiplot} show the rotation curves computed using standard {\it Newtonian} dynamics without dark matter for Sequence~1. As we can see, the velocities at radii larger than approximately 6\,kpc are larger in the polar plane. The reason for this is that the closed orbits are much more eccentric in the host galaxy than in the plane of the polar ring due to the compact host galaxy and the extended ring (see Fig.~\ref{fig:orbits}), and when observing the polar ring galaxy edge-on, the minimum velocities of these eccentric orbits (point A in Fig.~\ref{fig:orbits}) are measured.
The eccentricity can be explained as follows. The potential due to the compact host galaxy component appears nearly spherical at large\footnote{In the context of investigated PRG models, large means larger than the size of the host galaxy.} radii for test particles in both the plane of the host galaxy and the plane of the polar ring. The potential generated by the extended polar ring, however, does appear spherical to particles orbiting within the ring, but not to particles orbiting in the host plane.\footnote{We know this because in Newtonian dynamics the linearity of Poisson's equation allows us to separate and linearly add the different potentials.}
This gives rise to lower line-of-sight velocities when the host disk and polar rings are seen edge-on.

At radii smaller than the size of the hole (the region where the density increases with radius), the eccentricity argument turns around. The test particles are near the center of the galaxy and accordingly near the center of the hole of the polar ring. At these radii, all test particles experience the gravitational field caused by the polar ring rather spherically, while the potential by the galactic disk appears spherical only to the test objects orbiting within the disk, not to those moving in the polar plane.
The transition appears around approximately 6\,kpc, i.e. between the radial size of the galactic disk, which is smaller than 6\,kpc, and that of the polar ring, which is larger than 6\,kpc.

This ellipticity of the orbits in the host is enhanced by the actual presence of the hole at the center of the polar ring, because, in order to have the same mass in the polar structure as in a corresponding disk, one needs to increase the density at large $r$ in the polar plane, making it more extended (see Fig.~\ref{fig:ringmodel}).

\subsection{Milgromian dynamics}
\begin{figure*}
   \centering
   \includegraphics[width=17cm]{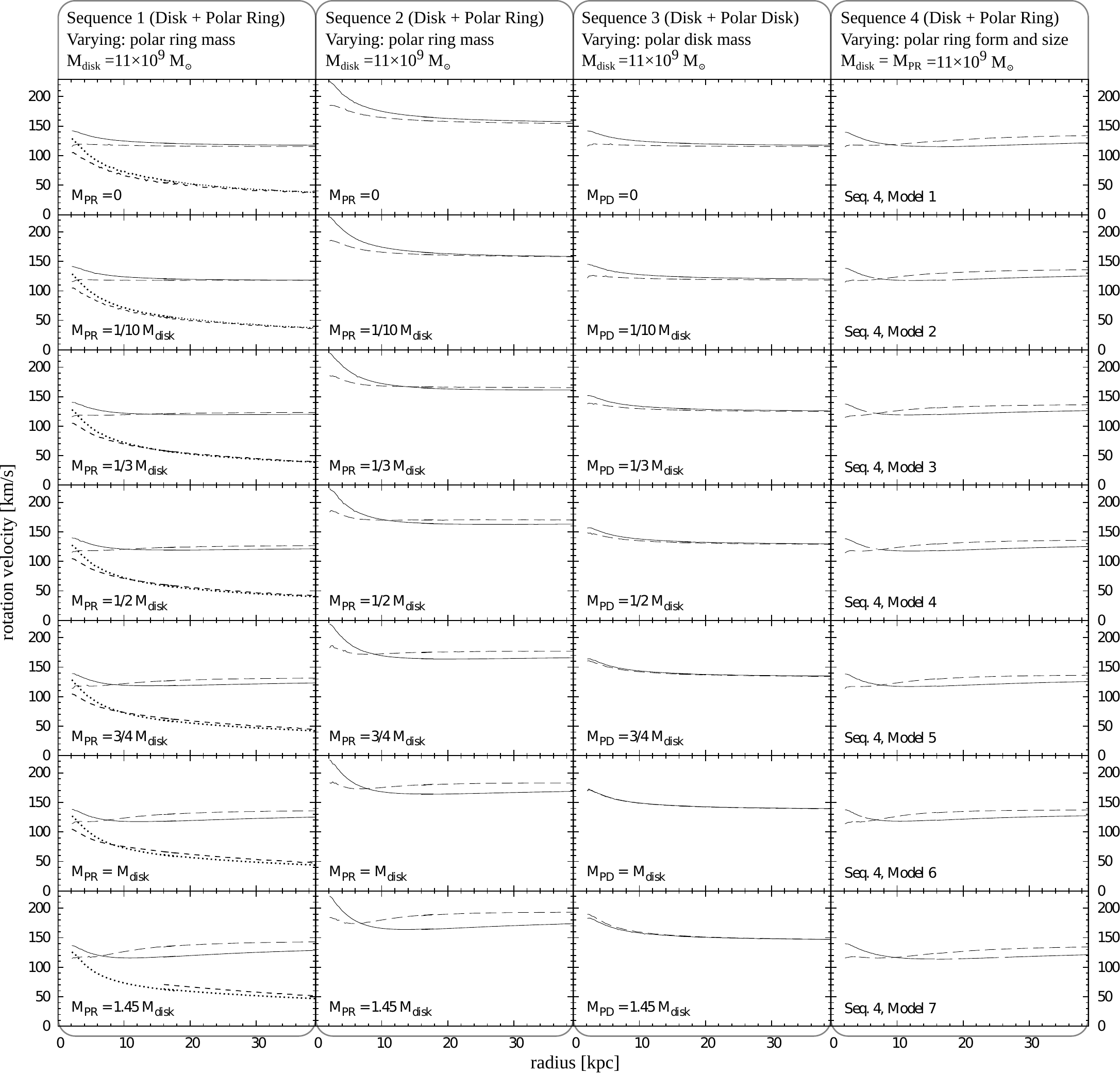}
      \caption{The figure presents the rotation curves of all models of Sequence~1~to~4. Solid lines refer to Milgromian rotation velocities in the host galaxy, long-dashed lines to the polar plane. In Seq.~1 also the Newtonian rotation curves are shown (dotted lines: host galaxy; short-dashed: polar ring; no dark matter halo).
      Sequence~1 features a host disk, a bulge and a polar ring. While the disk mass $M_\text{disk}=11\times10^9\,\Msun$ is constant, the mass of the polar ring is varied from $M_\text{PR} = 0 \dots 1.45\,M_\text{disk}$ (the bottom left panel shows the benchmark model, see also Fig.~\ref{fig:rotcurves}).
      Sequence~2 is similar to Sequence~1 but is three times as dense and consequently massive ($M_\text{disk} = 33\times 10^9\, \Msun$).
      Sequence~3 features a host disk, a bulge and a polar disk instead of a ring. Models of this sequence have a disk mass of $M_\text{disk}=11\times 10^9\,\Msun$, the mass of their polar component is again variable.
	Sequence~4 features a host disk, a bulge and a polar ring. Models of this sequence have a fixed total mass $M=27.1\times 10^9\,\Msun$ and the size parameters of their polar ring components are variable.}
	\label{fig:multiplot}
\end{figure*}

In Milgromian dynamics, all investigated models (see Fig.~\ref{fig:multiplot}) that (i) feature a polar ring (i.e. Sequences~1, 2, 4, 5 and 6), that (ii) has a total mass (i.e. gaseous plus stellar mass) comparable to the mass of the host galaxy (within a factor of $\sim$2), show higher velocities in the polar plane at radii larger than approximately 6\,kpc. Indeed, the reason is the same as in Newtonian dynamics without dark matter, and is even boosted by the additional gravity provided by Milgromian dynamics. Because the host is more compact than the ring, it appears more spherical to particles orbiting in the ring at large radii than the ring appears to particle orbiting in the host at the same radii. Hence, also in Milgromian dynamics, the closed orbits are in the galactic disk more eccentric than in the plane of the polar ring. This gives rise to lower line of sight velocities in the host for typical observed systems where the host and polar rings are seen approximately edge-on. Because Milgromian dynamics adds a disk component of PDM to the host and to the ring (see Fig.\ref{fig:phantom}), this effect is even amplified compared to Newtonian dynamics.

In Sequence~1 (Fig.~\ref{fig:multiplot}, 1st column), one can see that decreasing the polar ring mass compared to the host gradually cancels the above effect, because the gravity generated by the polar ring becomes more and more negligible when decreasing its mass. For the benchmark model, at a radius of less than approximately 6\,kpc, test stars orbiting in the polar ring have smaller velocities than those at the same radius in the host disk, because they are close to or in the polar hole and the mass enclosed by their orbits is comparably small. This transition radius changes to larger radii when decreasing the polar ring mass, to gradually arrive at the situation of no polar ring, where polar orbits have velocities systematically lower than those in the disk (equal at large radii).

The same effect is observed in Sequence~2 (Fig.~\ref{fig:multiplot}, second column) for more massive systems. Note that the velocities in the host decrease with declining polar ring mass, due to the decreasing gravity of the polar ring, but less so than the velocities in the ring. The reason for this slower decrease is that the effect of decreased gravity is compensated by the effect of decreasing eccentricity for particles orbiting in the host.

On the other hand, in Sequence~3 (Fig.~\ref{fig:multiplot}, 3rd column), the rotation velocities are, at radii larger than 15\,kpc, very similar in both planes, because of the special symmetry of these models. This emphasizes the role played by the hole at the center of the ring in the other Sequences.

Varying the form and size of the stellar and gaseous holes in the polar ring in Sequence~4 (Fig.~\ref{fig:multiplot}, 4th column) shows that for a host and ring of comparable mass, the effect is quite generic in the presence of a hole.

\begin{figure}
   \centering
   \includegraphics[width=8.5cm]{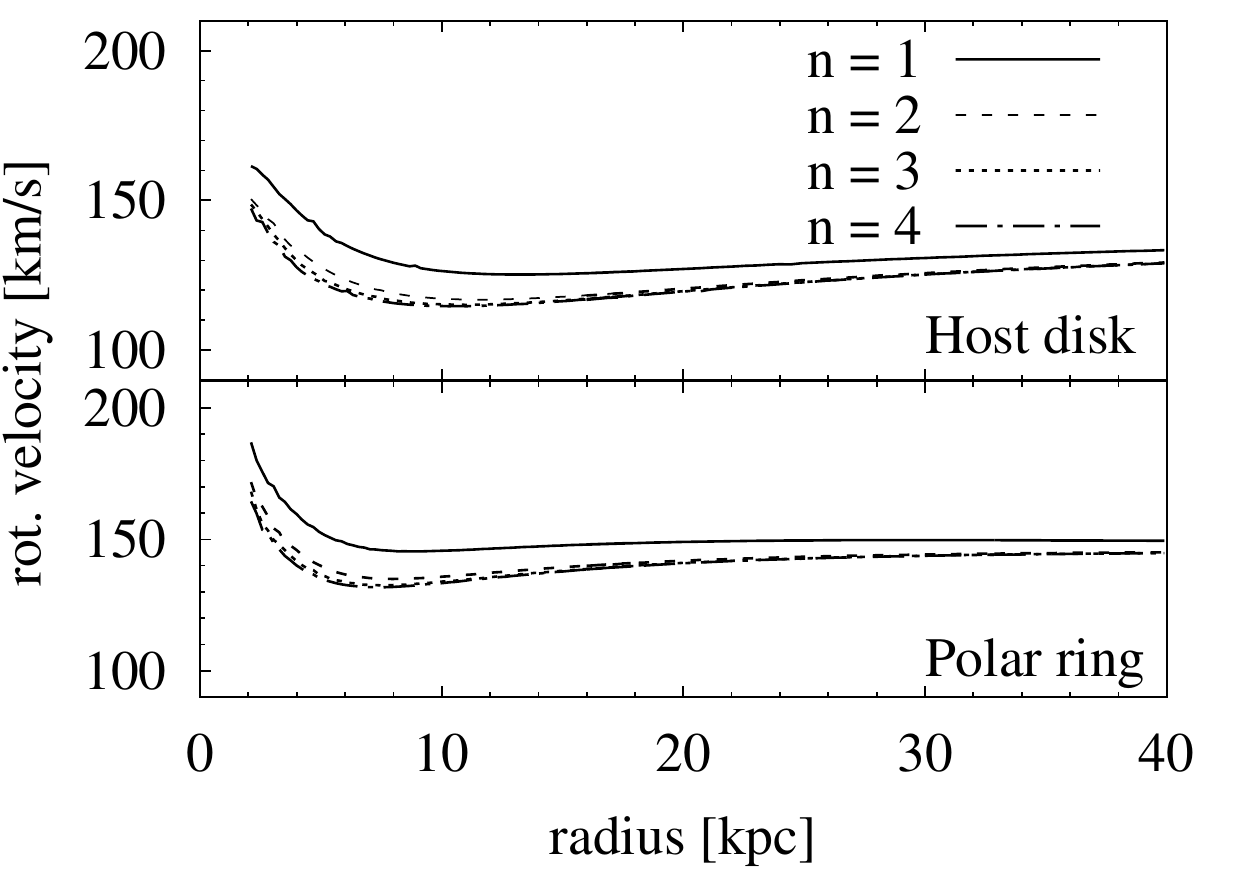}
      \caption{Rotation curves of the benchmark model in Milgromian dynamics applying different $\nu$-functions, $\nu_n(x)$ with $n=1,2,3,4$ in Eq.~\ref{eq:nu}.}
\label{fig:seq_nu}
\end{figure}

The shape of the rotation curves of Sequence~5 are all similar to the benchmark model, but their amplitude varies and is discussed in Sect.~6.3. Sequence~6 is a series of four different Milgromian potentials computed for the benchmark model for $n=1,2,3,4$ in Eq.~\ref{eq:nu}, and confirms that the qualitative results are independent of the $\nu$-function as illustrated on Fig.~\ref{fig:seq_nu}.

Let us finally note that if the host galaxy, here represented by a Miyamoto-Nagai disk, is replaced by a disk that falls off exponentially and therefore much faster, the host galaxy would appear more compact and the described effect would therefore be even stronger at small and intermediate radii, and unchanged at large radii.

\begin{figure}
   \centering
   \includegraphics[width=8.5cm]{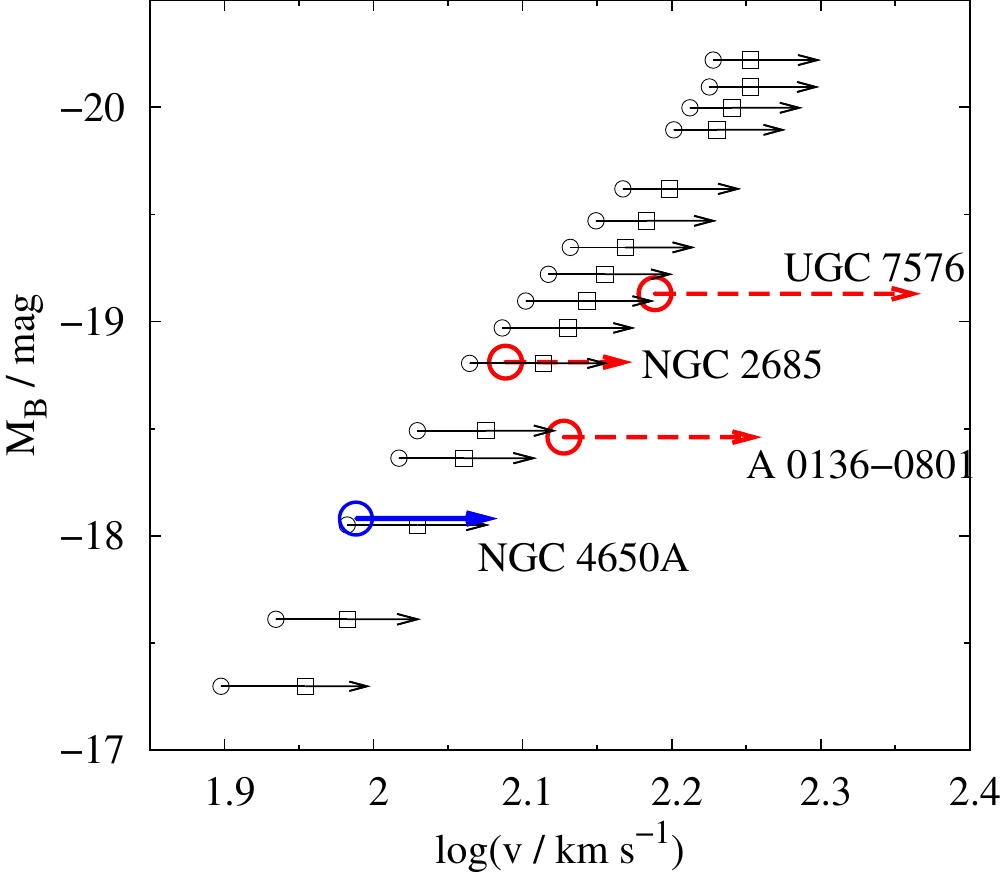}
      \caption{Comparison of observational data of PRGs with our numerical results using the luminous Tully-Fisher relation. The plot shows the the absolute B-band luminosity in magnitudes vs. the rotation velocity. Each arrow refers to one galaxy or galaxy model. The blue and red data points are adopted from \citet{Iodice03} and represent measurements of various PRGs. The circles show the rotation velocity measured in the hosts, the arrow heads the ones measured in the polar rings. The blue data correspond to NGC~4650A. For the theoretical data points (black), the squares show the rotation velocity in the host galaxy at $r=40\,\text{kpc}$, the circles show the rotation velocity in the host at $r=15\,\text{kpc}$ (where it is actually measured), the arrow heads point to the polar ring rotation velocity at $r=40\,\text{kpc}$. The theoretical data are obtained from models of Sequence~5. The absolute B-band magnitude is calculated from the total mass using a mass to light ratio of $M/L_\text{B}=4$ like was assumed by \citet{CombesArnaboldi}.}
\label{fig:tf}
\end{figure}

\subsection{Tully-Fisher relation}
We compare our theoretically obtained rotation curves to the observed PRGs of \citet{Iodice03}, showing, in Fig.~\ref{fig:tf}, the maximal rotational velocities of both the host and polar ring.
This gives rise to a luminous Tully-Fisher relation.
For the baryonic Tully-Fisher relation (BTFR), we note that Milgromian dynamics predicts $V^4 = a_0 G M_\text{b}$ for spherical systems \citep{StacyPRL} where $V$ is the asymptotic circular velocity, and $M_\text{b}$ the baryonic mass. However, polar ring galaxies are not only non-spherical but also non-axisymmetric objects. Because of this, there is no iron-clad prediction for the BTFR in such objects: our models indicate that the hosts typically exhibit asymptotic velocities below this value of $V$, while the polar rings are closer to the prediction, but can typically also exceed this velocity.

In order to explore the relevant mass range in our models, the corresponding data points in absolute B-band magnitude vs. rotation velocity are compared to the models of Sequence~5 (which varies the total mass), by assuming a mass to light ratio of $M/L_\text{B}=4$ as in \citet{CombesArnaboldi}. We assume that the line width $\Delta v_{20}=W_{20} = 2 v_\text{max}$ equals twice the maximum line of sight velocity\footnote{Note that the observed line widths $\Delta v_{20}$ are converted into velocities by assuming $\Delta v_{20} = 2v_\text{max}$ \citep[e.g.][]{Verheijen2001}. Depending on line width broadening effects, the actual velocities may be systematically smaller than $\Delta v_{20}/2$. In the context of the theoretical data this would imply that NGC~4650A had actually a smaller mass to light ratio.}.

From observations, the polar rings are generally very extended and feature large quantities of HI gas. The rotation curve in this polar plane can therefore observationally be measured even at large radii where it is very flat. Compared to the ring, the host galaxy is usually rather small and has relatively little gas. The rotation velocity in the host is generally obtained indirectly from the measured stellar velocity dispersion \citep{Iodice03,Iodice2006,Iodice2}, which means that the maximum velocity is most likely not measured in the very flat part and does therefore not equal the maximum velocity of the theoretical potential derived from the observed density distribution. To account for this issue, the theoretical rotation velocity from the models are computed at both $r=40\,\text{kpc}$ and $r=15\,\text{kpc}$, where $v(r=15\,\text{kpc}) \leq v(r=40\,\text{kpc})$ (see Fig.~\ref{fig:rotcurves}).

In Fig.~\ref{fig:tf}, the circles show the rotation velocity measured in the hosts, the arrow heads the ones measured in the polar rings, both for some observed PRGs and for the models of Sequence~5 (circles and squares as per Fig.~\ref{fig:rotcurves}). We see that our models reproduce fairly well the observations, with velocities systematically larger in the rings, but also with comparable offsets. Since the benchmark model on which the Sequences are based was inspired from the prototypical PRG galaxy NGC4650A, it comes as no surprise that this PRG is best fitted by these models.

\section{Conclusions}\label{sect:conclusions}
The conclusion from all the investigated models, and the bottom line of this study, is that Milgromian dynamics naturally predicts that rotation velocities would be higher in the polar rings than in the hosts. This generically happens when the ring is more extended than the host and of comparable mass and both are observed approximately edge-on. It does not apply to faint polar rings, or to polar rings of similar radial size as the host.
Given the wide range of model parameters covered within this study, this general result appears quite robust in Milgromian dynamics.
What is more, the magnitude of the velocity offset predicted by the models is also comparable to the observed one (see Fig.~\ref{fig:tf}). We however did not attempt to precisely fit the full rotation curves of individual polar ring galaxies, which will be the subject of future work, based on observations performed on a sample of 10 such systems at the Westerbork Synthesis Radio Telescope (WSRT)\footnote{PIs: G. Gentile and F. Combes}.
These and other upcoming precise measurements of rotation curves of individual polar ring galaxies \citep[e.g. the future measurements announced in][]{Iodice2} should thus allow more stringent tests, and these will become benchmark objects with which to test gravity in the coming years.

\begin{table*}
\begin{tabularx}{\textwidth}{cZZcccc}
	\hline
	Seq. &
	$M_\text{d}$ [$10^9\,\Msun$] &
	$M_\text{PR}/M_\text{d}$&
	$h_{\text{r}_1}^\text{gas}$ [kpc]  &
	$h_{\text{r}_2}^\text{gas}$ [kpc] &
	$h_{\text{r}_1}^\text{st}$ [kpc]  &
	$h_{\text{r}_2}^\text{st}$ [kpc] 
	\\
	\hline
	\\
	1	&	11.0	&	0.1, 0.25, 0.33, 0.5, 0.75, 1, 1.45  &  6.8	&  5.95	&  15.3	&  3.4 \\
	\\
	2	&	33.0	&	0.1, 0.25, 0.33, 0.5, 0.75, 1, 1.45  &  6.8	&  5.95	&  15.3	&  3.4 \\
	\\
	3	&	11.0	&	0.1, 0.25, 0.33, 0.5, 0.75, 1, 1.45  &  0.748	& -- 	&  0.748 & --  \\
	\\
	4	&	11.0	&	1.45 &  15.3	&  3.4  	&  15.3	&  3.4	\\
		&	&	&	7.8	&  4.95	&  15.3	&  3.4	\\
		&	&	&  5.8	&  4.95	&  15.3	&  3.4	\\
		&	&	&  6.8	&  5.95	&  15.3	&  3.4 \\
		&	&	&  6.8	&  5.95	&  11.3	&  7.4	\\
		&	&	&  6.8	&  5.95	&  6.8	&  5.95	\\
		&	&	&  10.8	&  9.95	&  10.8	&  9.95	\\
	\\
	5	& 2.8, 3.7, 5.6, 7.5, 8.4, 11.2, 13.0, 14.6, 16.4, 18.4, 20.7, 23.7, 30.6, 33.6, 36.8, 41.3  &  1.45 &  6.8		&  5.95	&  15.3	&  3.4 \\
	\hline
\end{tabularx}
\label{tab:ringsizes}
\caption{Model parameters. In the presented models of PRGs, each ring is built by the difference of two Miyamoto-Nagai disks (see the text for more details), whose radial parameters are $h_{\text{r}_1}$ and $h_{\text{r}_2}$. Model~7 of Seq.~1 and Model~4 of Seq.~4 equal the benchmark model. The models of Seq.~3 do not feature a polar ring but a polar disk of stars and gas with radial parameters $h_{\text{r}_1}^\text{gas/st}$. Seq.~6 is not included in this table, it corresponds to the benchmark model with 4 different $\nu$-functions.}
\end{table*}

\subsection*{Acknowledgements}
FL is supported by DFG grant \mbox{KR1635/16-1}. GWA's research is supported by the Claude Leon Foundation and a University Research Committee fellowship from the University of Cape Town. BF thanks the Humboldt Foundation for support at the beginning of this work. We thank an anonymous referee for very detailed comments that have improved the present manuscript.


\begin{thebibliography}{}

\bibitem[\protect\citeauthoryear{{Angus} \& {Diaferio}}{{Angus} \&
  {Diaferio}}{2011}]{Angus1}
{Angus} G.~W.,  {Diaferio} A.,  2011, \mnras, 417, 941

\bibitem[\protect\citeauthoryear{{Angus}, {van der Heyden}, {Famaey},
  {Gentile}, {McGaugh} \& {de Blok}}{{Angus} et~al.}{2012}]{Angus2}
{Angus} G.~W.,  {van der Heyden} K.~J.,  {Famaey} B.,  {Gentile} G.,  {McGaugh}
  S.~S.,    {de Blok} W.~J.~G.,  2012, \mnras, 421, 2598

\bibitem[\protect\citeauthoryear{{Bekenstein} \& {Milgrom}}{{Bekenstein} \&
  {Milgrom}}{1984}]{BM84}
{Bekenstein} J.,  {Milgrom} M.,  1984, \apj, 286, 7

\bibitem[\protect\citeauthoryear{{Bertone}, {Hooper} \& {Silk}}{{Bertone}
  et~al.}{2005}]{Bertone2005}
{Bertone} G.,  {Hooper} D.,    {Silk} J.,  2005, \physrep, 405, 279

\bibitem[\protect\citeauthoryear{{Brada} \& {Milgrom}}{{Brada} \&
  {Milgrom}}{1999}]{BradaMilgrom1999}
{Brada} R.,  {Milgrom} M.,  1999, \apj, 519, 590

\bibitem[\protect\citeauthoryear{{Combes} \& {Arnaboldi}}{{Combes} \&
  {Arnaboldi}}{1996}]{CombesArnaboldi}
{Combes} F.,  {Arnaboldi} M.,  1996, \aap, 305, 763

\bibitem[\protect\citeauthoryear{{Disney}, {Romano}, {Garcia-Appadoo}, {West},
  {Dalcanton} \& {Cortese}}{{Disney} et~al.}{2008}]{Disney08}
{Disney} M.~J.,  {Romano} J.~D.,  {Garcia-Appadoo} D.~A.,  {West} A.~A.,
  {Dalcanton} J.~J.,    {Cortese} L.,  2008, \nat, 455, 1082

\bibitem[\protect\citeauthoryear{{Famaey} \& {McGaugh}}{{Famaey} \&
  {McGaugh}}{2012}]{FamMcgaugh}
{Famaey} B.,  {McGaugh} S.~S.,  2012, Living Reviews in Relativity, 15, 10

\bibitem[\protect\citeauthoryear{{Gentile}, {Famaey} \& {de Blok}}{{Gentile}
  et~al.}{2011}]{things}
{Gentile} G.,  {Famaey} B.,    {de Blok} W.~J.~G.,  2011, \aap, 527, A76

\bibitem[\protect\citeauthoryear{{Iodice}}{{Iodice}}{2010}]{Iodice2}
{Iodice} E.,  2010, in {Debattista} V.~P.,  {Popescu} C.~C.,  eds, American
  Institute of Physics Conference Series Vol.~1240 of American Institute of
  Physics Conference Series, {Polar Disk Galaxies as New Way to Study Galaxy
  Formation: the Case of NGC4650A}.
pp 379--382

\bibitem[\protect\citeauthoryear{{Iodice}, {Arnaboldi}, {Bournaud}, {Combes},
  {Sparke}, {van Driel} \& {Capaccioli}}{{Iodice} et~al.}{2003}]{Iodice03}
{Iodice} E.,  {Arnaboldi} M.,  {Bournaud} F.,  {Combes} F.,  {Sparke} L.~S.,
  {van Driel} W.,    {Capaccioli} M.,  2003, \apj, 585, 730

\bibitem[\protect\citeauthoryear{{Iodice}, {Arnaboldi}, {Napolitano},
  {Oosterloo} \& {J{\'o}zsa}}{{Iodice} et~al.}{2008}]{Iodice4}
{Iodice} E.,  {Arnaboldi} M.,  {Napolitano} N.~R.,  {Oosterloo} T.~A.,
  {J{\'o}zsa} G.~I.~G.,  2008, in {Funes} J.~G.,  {Corsini} E.~M.,  eds,
  Formation and Evolution of Galaxy Disks Vol.~396 of Astronomical Society of
  the Pacific Conference Series, {Dark Matter Content in the Polar Disk Galaxy
  NGC 4650A}.
p.~483

\bibitem[\protect\citeauthoryear{{Iodice}, {Arnaboldi}, {Saglia}, {Sparke},
  {Gerhard}, {Gallagher}, {Combes}, {Bournaud}, {Capaccioli} \&
  {Freeman}}{{Iodice} et~al.}{2006}]{Iodice2006}
{Iodice} E.,  {Arnaboldi} M.,  {Saglia} R.~P.,  {Sparke} L.~S.,  {Gerhard} O.,
  {Gallagher} J.~S.,  {Combes} F.,  {Bournaud} F.,  {Capaccioli} M.,
  {Freeman} K.~C.,  2006, \apj, 643, 200

\bibitem[\protect\citeauthoryear{{Kroupa}}{{Kroupa}}{2012}]{Kroupa2}
{Kroupa} P.,  2012, \pasa, 29, 395

\bibitem[\protect\citeauthoryear{{Kroupa}, {Famaey}, {de Boer},
  {Dabringhausen}, {Pawlowski}, {Boily}, {Jerjen}, {Forbes}, {Hensler} \&
  {Metz}}{{Kroupa} et~al.}{2010}]{Kroupa1}
{Kroupa} P.,  {Famaey} B.,  {de Boer} K.~S.,  {Dabringhausen} J.,  {Pawlowski}
  M.~S.,  {Boily} C.~M.,  {Jerjen} H.,  {Forbes} D.,  {Hensler} G.,    {Metz}
  M.,  2010, \aap, 523, A32

\bibitem[\protect\citeauthoryear{{Kroupa}, {Pawlowski} \& {Milgrom}}{{Kroupa}
  et~al.}{2012}]{KroupaPawlowskiMilgrom2013}
{Kroupa} P.,  {Pawlowski} M.,    {Milgrom} M.,  2012, International Journal of
  Modern Physics D, 21, 30003

\bibitem[\protect\citeauthoryear{{McGaugh}}{{McGaugh}}{2011}]{StacyPRL}
{McGaugh} S.~S.,  2011, Physical Review Letters, 106, 121303

\bibitem[\protect\citeauthoryear{{Milgrom}}{{Milgrom}}{1983}]{Mil83}
{Milgrom} M.,  1983, \apj, 270, 365

\bibitem[\protect\citeauthoryear{{Milgrom}}{{Milgrom}}{2001}]{mondhalo}
{Milgrom} M.,  2001, MNRAS, 326, 1261

\bibitem[\protect\citeauthoryear{{Milgrom}}{{Milgrom}}{2009}]{BIMOND}
{Milgrom} M.,  2009, \prd, 80, 123536

\bibitem[\protect\citeauthoryear{{Milgrom}}{{Milgrom}}{2010}]{QUMOND}
{Milgrom} M.,  2010, \mnras, 403, 886

\bibitem[\protect\citeauthoryear{{Miyamoto} \& {Nagai}}{{Miyamoto} \&
  {Nagai}}{1975}]{MN}
{Miyamoto} M.,  {Nagai} R.,  1975, \pasj, 27, 533

\bibitem[\protect\citeauthoryear{{Moiseev}, {Smirnova}, {Smirnova} \&
  {Reshetnikov}}{{Moiseev} et~al.}{2011}]{Moiseev11}
{Moiseev} A.~V.,  {Smirnova} K.~I.,  {Smirnova} A.~A.,    {Reshetnikov} V.~P.,
  2011, \mnras, 418, 244

\bibitem[\protect\citeauthoryear{{Peebles} \& {Nusser}}{{Peebles} \&
  {Nusser}}{2010}]{Peebles}
{Peebles} P.~J.~E.,  {Nusser} A.,  2010, \nat, 465, 565

\bibitem[\protect\citeauthoryear{{Plummer}}{{Plummer}}{1911}]{plummer}
{Plummer} H.~C.,  1911, \mnras, 71, 460

\bibitem[\protect\citeauthoryear{{Reshetnikov} \& {Combes}}{{Reshetnikov} \&
  {Combes}}{1994}]{Reshetnikov1994}
{Reshetnikov} V.~P.,  {Combes} F.,  1994, \aap, 291, 57

\bibitem[\protect\citeauthoryear{{Sackett}, {Rix}, {Jarvis} \&
  {Freeman}}{{Sackett} et~al.}{1994}]{Sackett1994}
{Sackett} P.~D.,  {Rix} H.-W.,  {Jarvis} B.~J.,    {Freeman} K.~C.,  1994,
  \apj, 436, 629

\bibitem[\protect\citeauthoryear{{Sackett} \& {Sparke}}{{Sackett} \&
  {Sparke}}{1990}]{Sackett1990}
{Sackett} P.~D.,  {Sparke} L.~S.,  1990, \apj, 361, 408

\bibitem[\protect\citeauthoryear{{Schweizer}, {Whitmore} \&
  {Rubin}}{{Schweizer} et~al.}{1983}]{Schweizer1983}
{Schweizer} F.,  {Whitmore} B.~C.,    {Rubin} V.~C.,  1983, \aj, 88, 909

\bibitem[\protect\citeauthoryear{Strigari}{Strigari}{2012}]{Strigari2012}
Strigari L.~E., , 2012, arXiv:1211.7090

\bibitem[\protect\citeauthoryear{{Tiret} \& {Combes}}{{Tiret} \&
  {Combes}}{2007}]{Tiret2007}
{Tiret} O.,  {Combes} F.,  2007, \aap, 464, 517

\bibitem[\protect\citeauthoryear{{Verheijen}}{{Verheijen}}{2001}]{Verheijen2001}
{Verheijen} M.~A.~W.,  2001, \apj, 563, 694

\bibitem[\protect\citeauthoryear{{Whitmore}, {Lucas}, {McElroy},
  {Steiman-Cameron}, {Sackett} \& {Olling}}{{Whitmore}
  et~al.}{1990}]{Whitmore1990}
{Whitmore} B.~C.,  {Lucas} R.~A.,  {McElroy} D.~B.,  {Steiman-Cameron} T.~Y.,
  {Sackett} P.~D.,    {Olling} R.~P.,  1990, \aj, 100, 1489

\end{thebibliography}
\end{document}